\documentclass[aps,prb,showpacs,twocolumn,amssymb]{revtex4}

\usepackage{amsmath}
\usepackage{subfigure}
\usepackage{graphicx}

\input epsf

\def\prb{Phys. Rev. B }
\def\prl{Phys. Rev. Lett. }

\def\be{\begin{equation}}
\def\ee{\end{equation}}
\def\ba{\begin{eqnarray}}
\def\ea{\end{eqnarray}}
\def\ie{{\it i.e.} }

\def\etal{{\it et al.} }

\def\124{YBa$_2$Cu$_4$O$_8$ }

\def\C60{A$_x$C$_{60}$ }

\begin{document}

\title{A contractor-renormalization study of Hubbard plaquette clusters}

\author{Shirit~Baruch and Dror~Orgad}

\affiliation{Racah Institute of Physics, The Hebrew University, Jerusalem 91904, Israel}

\date{\today}

\begin{abstract}

We implement the contractor-renormalization method to study the
checkerboard Hubbard model on various finite-size clusters as
function of the inter-plaquette hopping $t'$ and the on-site
repulsion $U$ at low hole doping. We find that the pair-binding
energy and the spin gap exhibit a pronounced maximum at
intermediate values of $t'$ and $U$, thus indicating that
moderate inhomogeneity of the type considered here
substantially enhances the formation of hole pairs. The rise of
the pair-binding energy for $t'<t'_{\rm max}$ is kinetic-energy
driven and reflects the strong resonating valence bond
correlations in the ground state that facilitate the motion of
bound pairs as compared to single holes. Conversely, as $t'$ is
increased beyond $t'_{\rm max}$ antiferromagnetic magnons
proliferate and reduce the potential energy of unpaired holes
and with it the pairing strength. For the periodic clusters
that we study the estimated phase ordering temperature at
$t'=t'_{\rm max}$ is a factor of 2--6 smaller than the pairing
temperature.

\end{abstract}


\maketitle

\section{Introduction}
\label{intro}

It is by now generally accepted that spatial inhomogeneity may emerge either
as a static or as a fluctuating effect in strongly-coupled models of the high-temperature
superconductors, and indeed in many of the real materials.\cite{ourreview} What is far
from being settled is the issue of whether such inhomogeneity is {\it essential} to
the mechanism of high-temperature superconductivity from repulsive interactions. While most
researchers would probably answer this question in the negative one should bare in mind
the absence of a conclusive evidence that the single-band two-dimensional Hubbard model,
widely believed to be the "standard model" of high-temperature superconductivity, actually
supports superconductivity with a high transition temperature.\cite{aimi} On the other hand,
when examined on small clusters the same model and its strong-coupling descendent,
the $t-J$ model, exhibit robust signs of incipient superconductivity in the form of a
spin-gap and pair binding.\cite{ourreview} This fact points to the possibility that
the strong susceptibility towards pairing is a consequence of the confining
geometry itself.

\vspace{0.1cm}
This line of thought has been pursued in the past by considering the extreme
limit where the electronic density modulation is so strong that the system consists of
weakly coupled Hubbard ladders\cite{optimal-ladder,optimal-AFK} or plaquettes\cite{wf-steve}.
Beyond the questionable applicability of such models to the physical systems, which are at
most only moderately modulated, it is clear that strong inhomogeneity, even if beneficial
to pairing, is detrimental to the establishment of phase coherence and consequently to
superconductivity. On both counts it is, therefore, desirable to extend the analysis to
the regime of intermediate inhomogeneity.

\begin{figure}[t]
\includegraphics[angle=0,width=\linewidth,clip=true]{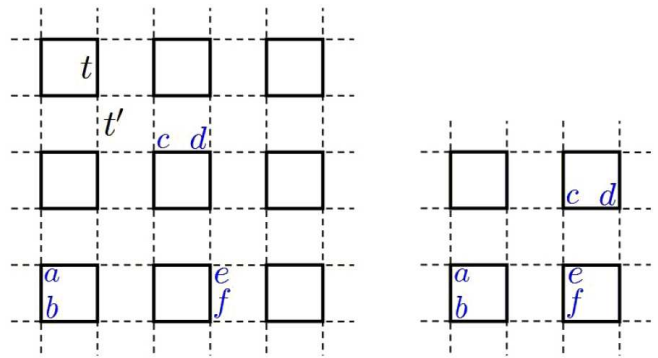}
\caption{The checkerboard Hubbard model. Shown here are two of the clusters that we
studied. The bonds labeled $ab$, $cd$, and $ef$ specify locations used in calculating the
pairing correlations.}
\label{model-fig}
\end{figure}

\vspace{0.15cm}
Recently, the checkerboard Hubbard model, constructed from 4-site
plaquettes with nearest-neighbor hopping $t$ and on-site repulsion $U$, was studied as
function of the inter-plaquette hopping $t'$ (see Fig. \ref{model-fig}).
Tsai \etal\cite{steve-exact} diagonalized exactly the $4\times 4$ site cluster
($2\times 2$ plaquettes) and found that the pair-binding energy, as defined by
Eq. (\ref{pb-def}) below, exhibits a substantial maximum at $t'\approx t/2$ for
$U\approx 8t$ and low hole concentration. Doluweera \etal\cite{DMFT-cluster},
on the other hand, used the dynamical cluster approximation in the range $0.8\le t'/t\le 1$
and obtained a monotonic increase in both the strength of the $d$-wave pairing interaction
and the superconducting transition temperature, $T_c$, towards a maximum that occurs in
the homogeneous model.

\vspace{0.1cm}
In this paper, we use the contractor-renormalization (CORE) method\cite{CORE} to
derive an effective low-energy Hamiltonian for the checkerboard Hubbard model, which we
then diagonalize numerically on various finite-size clusters. We begin by establishing the
region of applicability of the CORE approximation by contrasting its predictions with the
exact results of Ref. \onlinecite{steve-exact} for $2\times 2$ plaquettes. Our findings
indicate that at low concentrations of doped holes the two approaches agree reasonably well
unless $t'$ is larger than a value, which increases with $U$. Deviations also appear for
small $t'$ when $U$ is large. We identify probable sources of these discrepancies.

\vspace{0.1cm}
Based on the lessons gained from the small system we go on to study larger clusters
of up to 10 plaquettes. These include the periodic $6\times 6$ sites cluster and 2-leg
and 4-leg ladders with periodic boundary conditions along their length.
Within the region where CORE is expected to provide reliable
results the pair-binding energy continues to exhibit a non-monotonic behavior with
a pronounced maximum at intermediate values of $t'$ and $U$. The precise location of
the maximum depends on the cluster geometry but it typically occurs in the range
$t'_{\rm max}\approx 0.5-0.7t$ and $U_{\rm max}\approx 5-8t$. The spin gap of the doped
system follows a similar trend, often reaching the maximum slightly before the pair-binding
energy. These findings demonstrate that moderate inhomogeneity, of the type
considered here, can substantially enhance the binding of holes into pairs.

\vspace{0.1cm}
In an effort to elucidate the source of the maximum we have looked into the content
of the ground state and calculated the contributions of various couplings in the
effective Hamiltonian to its energy. Our results indicate that for $t'<t'_{\rm max}$
the doped holes move in a background, which is composed predominantly of plaquettes
that are in their half-filled ground state. This background possesses strong intra-plaquette
singlet resonating valence bond (RVB) correlations, which facilitate the propagation
of pairs relative to independent holes. The rise in the pair-binding energy
while $t'$ grows towards $t'_{\rm max}$ is a result of a faster decrease
of the pair kinetic energy in comparison to that of unpaired fermions. As $t'$
crosses $t'_{\rm max}$ and approaches the uniform limit the ground state contains
a growing number of plaquettes that support antiferromagnetic (AFM) magnons. In this
regime of increasing AFM correlations the kinetic energy changes relatively little with
$t'$, and the decrease of the pair-binding energy for $t'>t'_{\rm max}$ is caused by
the lowering of the energy of single holes due to their interactions with the magnons.
Interestingly, we find that the maximum in the pair-binding energy of the periodic clusters
is accompanied by a change in the crystal momentum of the single-hole ground state from
the $\Gamma-{\rm M}$ and symmetry related directions at $t'<t'_{\rm max}$ to the
Brillouin-zone diagonals at $t'>t'_{\rm max}$. A similar correlation was also
found for the 3-hole ground state of the $6\times 6$ sites cluster.

While the pair-binding energy sets a pairing scale, $T_p$, a phase-ordering scale, $T_\theta$,
is provided by the phase stiffness. The latter was evaluated from the second derivative of
the ground state energy with respect to a phase twist introduced by threading the system with
an Aharonov-Bohm flux. We have found that as the twist is taken to zero, the CORE energy
curvature typically converges towards a limiting value only when $t'<t'_{\rm max}$.
Within this region the phase stiffness increases monotonically with $t'$.
Our results indicate that for the lightly doped periodic clusters that we have considered
phase fluctuations dominate over pairing, specifically, $T_p\approx 2-6 T_\theta$
at $t'=t'_{\rm max}$. The limitations of the present study
make it difficult to draw conclusions regarding the behavior of $T_c$ in the
two-dimensional thermodynamic limit.

We have also calculated the pair-field correlations between Cooper-pairs that reside on the
most distant bonds allowed by our finite clusters. As expected, these correlations are
consistent with $d$-wave pairing. However, in contrast to the pair-binding energy and
the phase stiffness the
correlations change little with $t'$ and are small in magnitude. This discrepancy
might be resolved in light of our finding that only few holes are tightly bound into pairs that
reside within a single plaquette. Moreover, we obtain that the number of such pairs changes
relatively little with $t'$ with no apparent correlation to the substantial maximum in the
pair-binding energy. Taken together these findings suggest that the correlation function
which we and others often use to identify and quantify pairing in the Hubbard model may be
ill-constructed to take account of the more extended and structured nature of
pairing in this model.

\section{Model and Method}
\label{models}

The Hamiltonian of the checkerboard Hubbard model, which we have studied, is given by
\be
H=-\sum_{\langle i,j\rangle , \sigma}\left( t_{ij} c_{i,\sigma}^\dagger c_{j,\sigma}
+{\rm H.c.}\right)+U\sum_i n_{i,\uparrow} n_{i,\downarrow},
\label{H}
\ee
where $c_{i,\sigma}^\dagger$ creates an electron with spin $\sigma=\uparrow,\downarrow$
at site $i$ of a two-dimensional square lattice. Here
$n_{i,\sigma}=c_{i,\sigma}^\dagger c_{i,\sigma}$, and $\langle i,j\rangle$ denotes
nearest-neighbor sites. The hopping amplitude is $t_{ij}=t$ for $i$ and $j$ on the same
plaquette, while $t_{ij}=t'$ when they belong to neighboring plaquettes, as shown in
Fig. \ref{model-fig}.

The first step in obtaining the CORE effective Hamiltonian for the above model,
is the exact diagonalization of a four-site plaquette. Out of the full spectrum, the $M$
lowest-energy states are retained. The reduced Hilbert space, in which the effective
Hamiltonian operates, is spanned by the tensor products of these states on different
plaquettes. Next, the Hamiltonian (\ref{H}) is diagonalized on $N$ connected plaquettes
and the $M^N$ lowest-energy states are projected onto the reduced Hilbert space and
Gram-Schmidt orthonormalized. Finally, after replacing the exact eigenstates by their
projections, the $N$-plaquette Hamiltonian can be represented as one for $M$ types of
hard core particles coupled via $N$-body interactions. The CORE approximation consists
of applying the resulting effective Hamiltonian to the study of larger clusters. By
construction, the spectrum of the CORE Hamiltonian coincides with the low-energy
spectrum of the exact problem on $N$ plaquettes. We note, however, that this ceases
to be the case if one or more of the exact low-energy states have zero projection
on the reduced Hilbert space, or, if some of them are projected onto the same
tensor-product state. In the following we demonstrate that such a problem arises in
certain parameter regions of the model (\ref{H}).

We concentrate on relatively low hole densities as measured from the half-filled
system. The simplest truncation used to describe this regime is to retain the ground
state of the half-filled plaquette [a total spin singlet $S=0$ with plaquette
momentum ${\bf q}=(0,0)$], its $S=1$, ${\bf q}=(\pi,\pi)$ triplet of lowest lying AFM
magnon excitations, and the $S=0$, ${\bf q}=(0,0)$ hole pair ground state.\cite{AAcore}
The inclusion of the magnon excitations is essential for retrieving the correct magnetic
behavior at low hole doping.\cite{core-res} Below we show that they also play an important
role in the physics of hole binding. One can improve the approximation by including in
the CORE plaquette basis also the two degenerate doublets $S_z=\pm 1/2$,
${\bf q}=(0,\pi),(\pi,0)$, comprising the single hole ground state.\cite{core-res,core-tj}
Moreover, the inclusion of these states is mandatory for the purpose of calculating
the pair binding energy, which is one of the goals of the present work. Consequently,
our CORE scheme consists of keeping the above mentioned $M=9$ states. We have considered
only range-2 interactions, \ie $N=2$.

The resulting effective Hamiltonian includes all possible couplings, which respect the
symmetries of the 2-plaquettes problem. These include the conservation of number of holes
$N_h$, invariance under SU(2) spin rotations and under reflections about the central
bonds of the cluster in the $x$ and $y$ directions. The latter, together with the conservation
of $N_h$, imply that within our reduced Hilbert space, as defined above, the total plaquette
momentum ${\bf q_1+q_2}$ is also conserved (modulo $2\pi$). We will not list here the 45
couplings which are allowed by the symmetries. Instead, we will describe the most important
ones in the appropriate context and refer the reader to the Appendix for a detailed
description of the Hamiltonian.

Many of the results reported in the following are derived from the spectrum of the
effective Hamiltonian as obtained by exact diagonalization.
We also calculate various
ground-state correlations. To this end we project the appropriate operators on the
reduced Hilbert space\cite{CORE} before evaluating their ground-state correlation
function.

\section{Results}
\label{results}

Although the size of the Hilbert space is massively reduced by the CORE approximation it
still grows exponentially with the size of the system. Therefore, even the largest clusters
that we are able to diagonalize using this method are too small for a direct calculation of
$T_c$. Instead we calculate various properties of the system which are indicative of the
two necessary ingredients for superconductivity: pairing and phase stiffness. We begin with
the former and study its behavior as function of $t'$ and $U$ on various geometries. These
include the $4\times 4$ and $6\times 6$ periodic clusters, seen in Fig. \ref{model-fig},
as well as 2-leg and 4-leg ladders with periodic boundary conditions along their length,
which extends up to 20 sites.

\begin{figure}[ht]
\includegraphics[angle=0,width=\linewidth,clip=true]{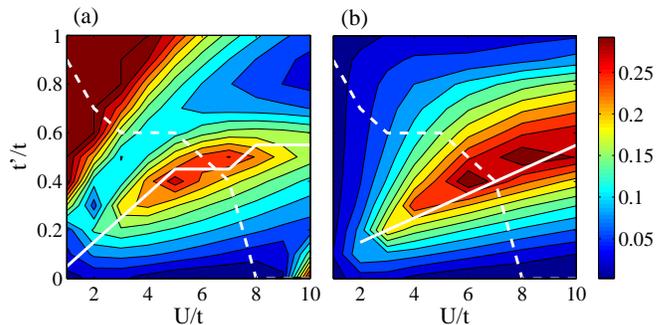}
\caption{The pair-binding energy in a periodic $4\times 4$ cluster at 1/16 hole doping
as obtained by (a) CORE, and (b) exact diagonalization (Ref. \onlinecite{steve-exact}).
CORE projects out low energy states from the effective Hilbert space in the region above
the dashed line. The crystal momentum of the degenerate single-hole ground state
is $(0,\pi)$ and $(\pi,0)$ below the solid line and $(0,0)$ and $(\pi,\pi)$ above it.}
\label{bindingCont1}
\end{figure}
\begin{figure}[ht]
\includegraphics[angle=0,width=\linewidth,clip=true]{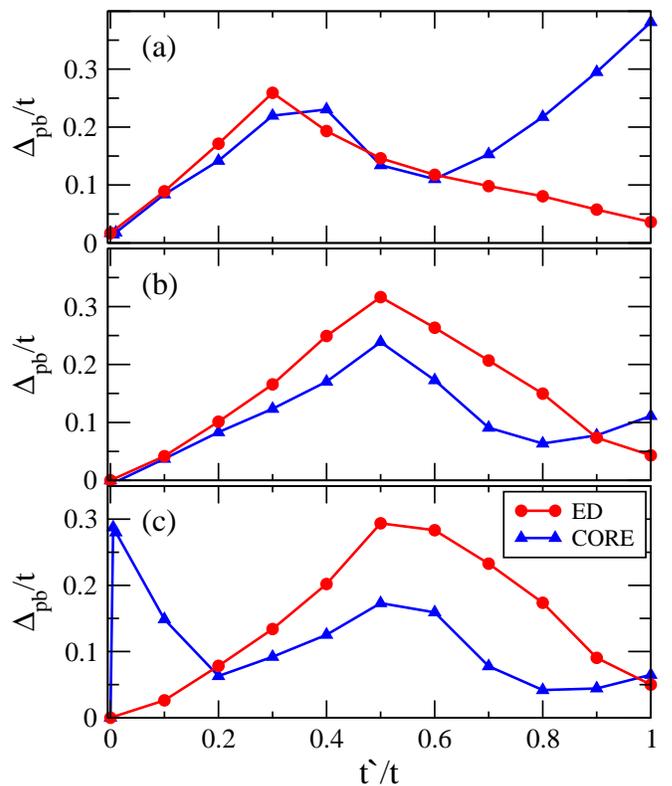}
\caption{The pair-binding energy in a periodic $4\times 4$
cluster at 1/16 doping for various values of the interaction strength (a) $U=4t$,
(b) $U=8t$, and (c) $U=10t$. Triangles depict the CORE results and circles
correspond to the exact diagonalization results of Ref. \onlinecite{steve-exact}.}
\label{binding1}
\end{figure}

\subsection{Pair-binding energy and spin-gap}

The pair-binding energy is defined by
\be
\Delta_{pb}(M/N)=2E_0(M)-\left[E_0(M+1)+E_0(M-1)\right],
\label{pb-def}
\ee
where $E_0(M)$ is the ground-state energy of the system with $M$ holes doped into the
$N$-site half-filled cluster. Consider two identical clusters each with $M$ holes.
If holes tend to pair and $M$ is odd it should be energetically favorable to move
an electron from one cluster to another in order to obtain a fully-paired state in both.
On the other hand, such a redistribution should be unfavorable if $M$ is even. In this sense,
a positive $\Delta_{pb}$ for odd $M$ and a negative $\Delta_{pb}$ for even $M$ signifies
an effective attraction between holes.

Recently, Tsai \etal\cite{steve-exact} have found by exact diagonalization of the periodic
$4\times 4$ cluster that the pair-binding energy exhibits a pronounced maximum both as
function of $t'$ and $U$. Their results allow for a critical evaluation of the validity of
the CORE method in a range of parameters. To this end we present in Figs. \ref{bindingCont1}
and \ref{binding1} a comparison between the CORE and the exact results for $\Delta_{pb}(1/16)$.
It is clear that CORE introduces substantial errors in two specific regimes: small $U$ and
large $t'$ [Fig. \ref{binding1}(a)], and large $U$ and small $t'$ [Fig. \ref{binding1}(c)],
while it is in reasonable agreement with the exact results in the intermediate parameter regime.

An obvious source for the discrepancies is the fact that our CORE approximation includes
only range-2 couplings. Longer-range interactions are expected to become more important
as the system becomes more homogeneous \ie when $t'\rightarrow t$.
We believe that the deviations between the CORE predictions and the exact results in this limit,
especially for small $U$ where the pair size is expected to be large, are
mainly due to insufficient range of the effective interactions.
A related problem may emerge at large $U$ where the extent of magnetic correlations grow.
However, we did not confirm these conjectures by explicit calculations.

\begin{figure}[ht]
\includegraphics[angle=0,width=\linewidth,clip=true]{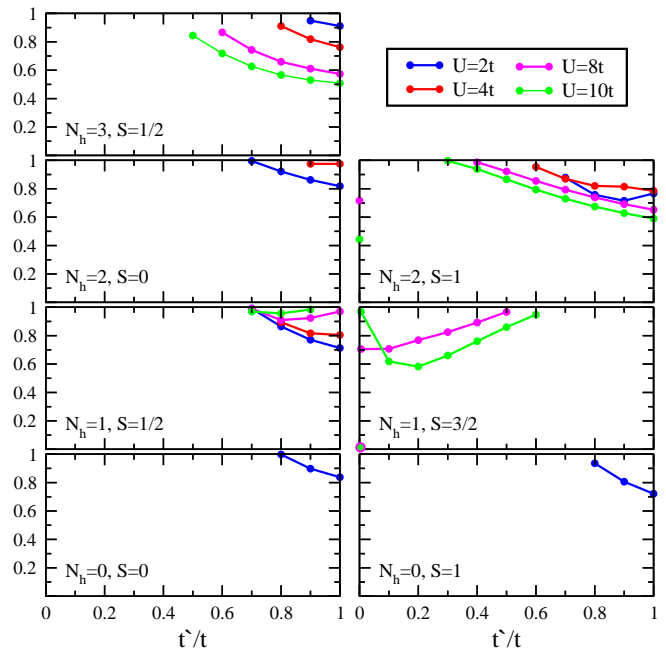}
\caption{The excitation energy of the lowest energy state that is projected-out by CORE
in units of the bandwidth of the kept states in its sector.}
\label{p-out}
\end{figure}

\begin{figure}[t!]
\includegraphics[angle=0,width=\linewidth,clip=true]{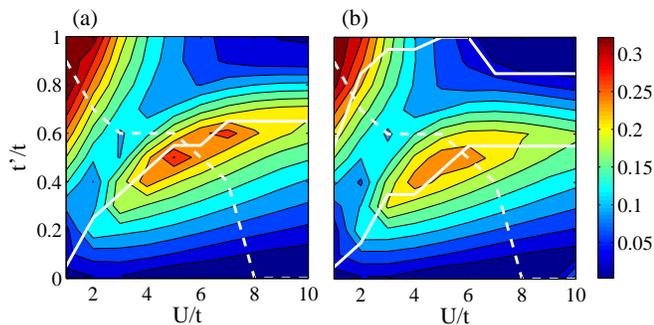}
\caption{The pair-binding energy in a periodic $6\times 6$ cluster at (a) 1/36, and (b) 3/36
hole doping. CORE projects out low energy states from the effective Hilbert space in the
region above the dashed line. In (a) the crystal momentum of the degenerate single-hole
ground state is $(0,\pm2\pi/3)$ and $(\pm2\pi/3,0)$ below the solid line and
$(\pm2\pi/3,\pm2\pi/3)$ above it. In (b) the crystal momentum of the degenerate 3-hole
ground state is $(\pm2\pi/3,\pm2\pi/3)$ between the solid lines and $(0,\pm2\pi/3)$ and
$(\pm2\pi/3,0)$ elsewhere.}
\label{bindingCont2}
\end{figure}

A more subtle source of errors, which we have mentioned already in the previous Section,
is the fact that low-energy states may be projected out from the CORE effective Hilbert
space in the process of generating the effective Hamiltonian. This happens when a low-lying
state of a connected cluster has zero overlap with the tensor-product states of the
effective Hilbert space or when two or more low-lying states are mapped onto the same state
in the effective space (Note, however, that spin-rotation symmetry is preserved in the
sense that spin multiplets are either kept or projected out as a whole.)
Fig. \ref{p-out} depicts for each of the sectors in which such a problem arises the excitation
energy of the lowest projected-out state in units of the bandwidth of the kept states in
the sector. We also denoted in Figs. \ref{bindingCont1} and \ref{bindingCont2}
the parameter region where the problem occurs.

The overlap issue is responsible for the failure of CORE in the regime of small $t'$ and
large $U$. When $U>7.858t$ and for $t'=0$ the $N_h=1$, $S=3/2$ double-plaquette (eight-fold
degenerate) ground state $|N_h=1,S=3/2\rangle_2$ consists of one plaquette in its half-filled
ground state and a second plaquette in a fully-polarized $S=3/2$ single-hole state. The latter
resides outside the effective Hilbert space and therefore $|N_h=1,S=3/2\rangle_2$ is projected
out. This ceases to be the case once $t'$ is turned on as a result of a component which
appears in $|N_h=1,S=3/2\rangle_2$ and corresponds to a system with a magnon on one
plaquette and a plaquette-fermion on the other. However, the amplitude of this component
diminishes with increasing $U$. This leads CORE to misidentify
the nature of $|N_h=1,S=3/2\rangle_2$ and induces an abrupt increase in the
magnon-fermion interaction [$V_{ft}^{3/2,\nu,{\bf q}}$ in Eq. (\ref{Vt})] for small $t'$.
As a result, CORE underestimates the energy of the two-hole ground state of the $4\times 4$
cluster and consequently predicts an erroneously large pair-binding energy, see
Fig. \ref{binding1}. Nevertheless, it appears that away from this region of parameters
the projected-out states are high enough in energy as to not cause qualitative errors.

\begin{figure}[th!]
\includegraphics[angle=0,width=\linewidth,clip=true]{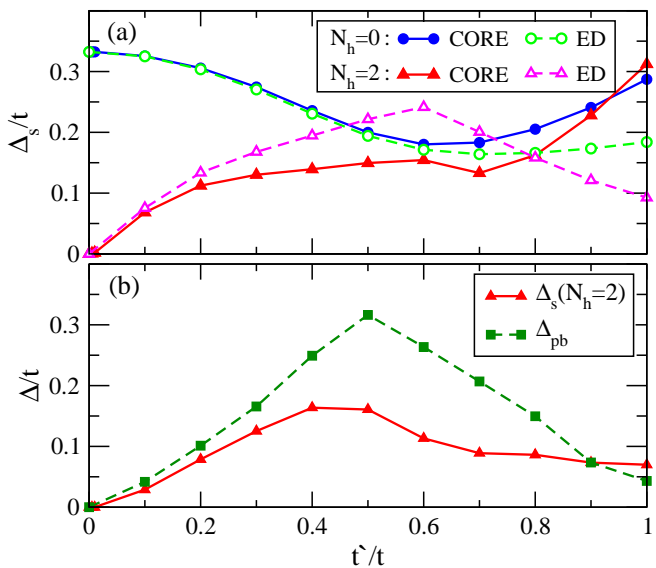}
\caption{The spin gap of undoped and two-hole doped systems at $U=8t$.
(a) The $4\times 4$ periodic cluster - comparison between CORE and exact diagonalization results.
(b) CORE results for the spin-gap $\Delta_{\text {\small{\it s}}}$ and the pair-binding energy
$\Delta_{pb}(1/36)$ of the $6\times 6$ periodic cluster.}
\label{spinGap}
\end{figure}

Based on the comparison of $\Delta_{pb}$ depicted in Figs. \ref{bindingCont1},\ref{binding1}
and similar plots presented below for the spin-gap [Fig. \ref{spinGap}(a)] and pair-field
correlations [Fig. \ref{pairingCorr}(a)] we conclude that CORE agrees semi-quantitatively with
the exact results provided $U/50 \lesssim t'\lesssim U/8$. Within this region, and across all
geometries studied, we found the pair-binding energy to exhibit the same qualitative behavior
consisting of a broad peak both as function of $t'$ and $U$. This conclusion holds true also
when one varies the doping level (at least in the low-doping regime which we have considered)
as can be seen from the results for the $6\times 6$ cluster presented in Fig. \ref{bindingCont2}.
In addition, the same figure suggests that the above mentioned problems with the CORE
method become less severe as the size of the system increases.


\begin{figure}[t!!!]
\includegraphics[angle=0,width=\linewidth,clip=true]{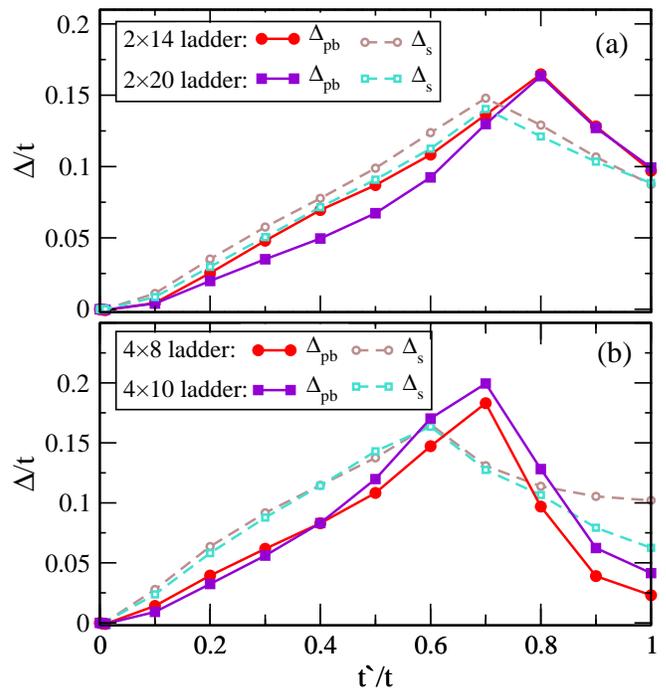}
\caption{The spin-gap $\Delta_{\text {\small{\it s}}}$ and pair-binding energy
$\Delta_{pb}$ at $U=8t$ for two-hole
doped a) 2-leg ladders,  (b) 4-leg ladders.}
\label{binding2}
\end{figure}

The association of positive pair-binding energy with Cooper pairing may be contested on the
ground that it can also be taken as evidence for a tendency of the system to phase separate.
We believe that this is not the case for the model studied here for the following reasons.
First, in accordance with the interpretation discussed above of $\Delta_{pb}$ as indication
for hole pairing we have found its sign to change according to $(-1)^{M+1}$ for all the clusters
and doping levels which we have considered. Second, while the appropriate criteria for identifying
regimes of phase separation from finite size studies include the Maxwell
construction\cite{Hellberg} and measurements of the surface tension in the presence of
boundary conditions that force phase coexistence, a crude way of identifying phase separation
is by calculating the inverse compressibility $\kappa^{-1}=n^2 \partial\mu/\partial n$, where
$\mu$ is the chemical potential and $n$ the electronic density. For numerical purposes a discrete
version is used, which in our case reads
\be
\kappa^{-1}\propto E_0(M+2)+E_0(M-2)-2E_0(M).
\label{comp-def}
\ee
Negative inverse compressibility indicates instability towards phase separation. We always find
$\kappa^{-1}>0$. Finally, whenever the ground state is a spin singlet one can define the spin-gap
as the energy gap to the lowest $S=1$ excitation. We have calculated the spin gap for the two-hole
doped systems and found that in all cases it follows the pair-binding energy in the regime of
small to moderate $t'$, see Figs. \ref{spinGap} and \ref{binding2}. This coincidence strongly
suggests that in this regime the lowest $S=1$ excitation is a result of a dissociation of a
hole pair into two separate holes. It is interesting to note that we always observe that the
spin-gap reaches a maximum and starts to drop before the pair-binding energy does so. This may
be an indication that moderate inhomogeneity supports the formation of a bound
$S=1$ magnon--hole-pair state.\cite{core-res,spingap2leg}

Consequently, our findings and the above arguments lead us to conclude that inhomogeneity of the
type included in the checkerboard Hubbard model substantially enhances hole-pairing. The precise
position of the point of optimal inhomogeneity in the sense of strongest pairing depends on the
cluster geometry and interaction strength. Albeit, it typically occurs
in the range $t'_{\rm max}\approx 0.5-0.7t$ and $U_{\rm max}\approx 5-8t$. We note that this
fact implies that the physics behind the large pairing scale of the model necessarily involves
inter-plaquette couplings since the single plaquette does not support hole-pairing beyond
$U_c\approx 4.6t$.\cite{AAcore}

\subsection{Energetics and structure of the ground state}

\begin{figure}[t]
\includegraphics[angle=0,width=\linewidth,clip=true]{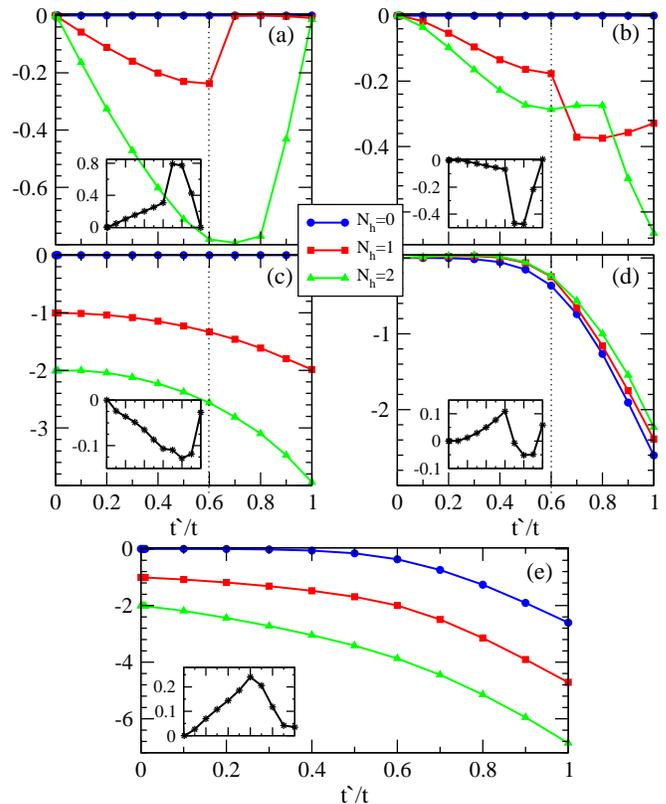}
\caption{Ground state expectation values of various effective couplings for the
$6\times 6$ periodic cluster at $U=8t$: (a) fermion and pair hopping; (b) fermion and pair
magnon-assisted hopping; (c) fermion and pair interactions; (d) interactions involving magnons;
(e) the full Hamiltonian. The insets show the contribution of each group of couplings to the
pair-binding energy. The full binding energy reaches a maximum at $t'=0.6t$ as indicated by
the dotted line.}
\label{coupling1}
\end{figure}

What drives the enhancement of hole-pairing and what is the reason for its maximum
as function of $t'$? In an attempt to gain insights into these questions we have took
advantage of the fact that CORE provides us with an effective Hamiltonian whose various
couplings can be classified and analyzed. To this end we have divided the 45 different
couplings into four groups, as described in the Appendix. They include: fermion and hole-pair
"bare" kinetic terms (including fermion and pair hopping as well as Andreev-like pair creation
and disintegration), magnon-assisted fermion and pair hopping, fermion and pair interactions
and finally, interactions involving magnons. Fig. \ref{coupling1} depicts the contribution
of each group to the ground-state energy of the $N_h=0,1,2$ doped $6\times 6$ periodic
cluster and to its pair-binding energy $\Delta_{pb}(1/36)$ at $U=8t$.

Fig. \ref{coupling1} makes it clear that the increase in the pair-binding energy from $t'=0$
to $t'_{\rm max}$ is dominated by a faster decrease of the kinetic energy of hole pairs as
compared to unpaired holes. Furthermore, in this region the pair-binding energy is largely
determined by the "bare" kinetic terms while the (negative) contribution of hopping processes
that involve magnons is much smaller. The small contributions of the various interactions
approximately cancel out. Looking more closely at the way charges propagate in this range of
$t'$ we found that the main channel for single holes is a direct hop between neighboring
plaquettes but that this process is virtually non-existent for hole pairs. Instead, a pair
propagates predominantly by Andreev-like dissociation into single holes on adjacent
plaquettes and recombination of these holes into a pair one register away from its original
position [as described by the last term in Eq. (\ref{Kbf})].

For $t'>t'_{\rm max}$ the behavior changes qualitatively and rather abruptly. The gain in
kinetic energy of the pair relative to that of unpaired holes ceases to increase. While
pairs continue to propagate mainly via a series of dissociation and recombination events,
single holes move almost exclusively by hopping processes involving magnons [the second and
third terms in Eq. (\ref{Kbft})]. The decrease in the pair-binding energy in this regime is
induced by a sharp decrease of the potential energy of the unpaired holes owing to their
interactions with the magnons. On the other hand, the contribution of interactions not
involving the magnons to the pair-binding energy does not show a significant change as
$t'$ is driven through $t'_{\rm max}$.

\begin{figure}[t]
\includegraphics[angle=0,width=\linewidth,clip=true]{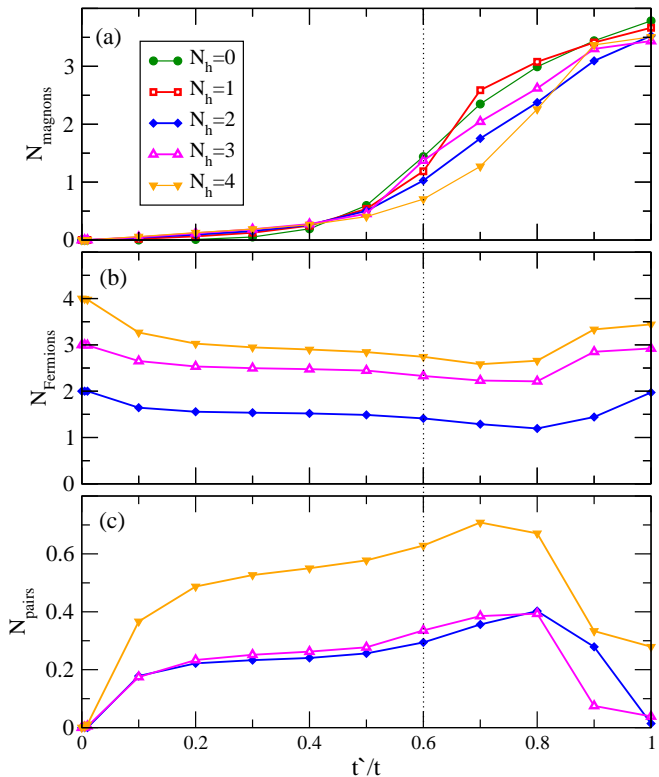}
\caption{ The average number of (a) magnons, (b) fermions, and (c) hole pairs in the
ground state of the $6\times 6$ periodic cluster at $U=8t$. The position of the pair-binding
energy maximum is indicated by the dotted line.}
\label{eigens2}
\end{figure}

\begin{figure}[t]
\includegraphics[angle=0,width=\linewidth,clip=true]{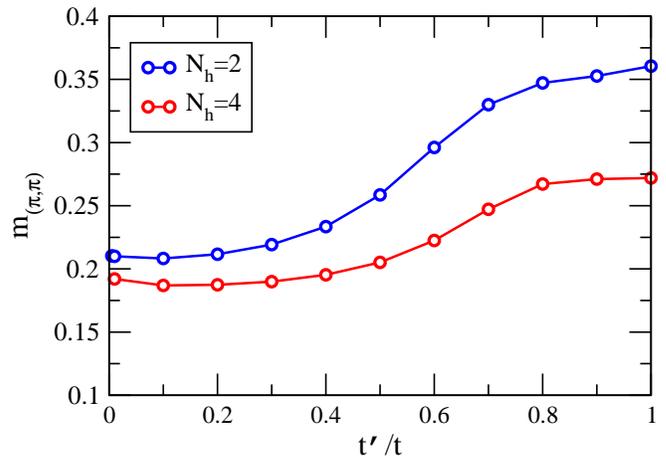}
\caption{The staggered magnetization in the two-hole and four-hole ground states of the
$6\times 6$ periodic cluster at $U=8t$.}
\label{magnetization}
\end{figure}

The above results suggest that the AFM magnons play an important role in inducing the
change in the behavior of the pair-binding energy. To further test this conclusion we have
looked at the evolution of the ground-state content with $t'$. Fig. \ref{eigens2} shows
the average number of magnons, fermions and pairs in the $N_h=0-4$ ground states of the
$6\times 6$ periodic cluster. Evidently, the magnons begin to proliferate slightly before
the maximum in the pair-binding energy is reached. Concomitantly, there is an increase
of AFM correlations in the system as can be seen from Fig. \ref{magnetization}, which
depicts  the staggered magnetization $m_{(\pi,\pi)}$ defined by
\be
m_{(\pi,\pi)}^2=\left\langle\left[\frac{1}{N}\sum_{j=1}^N e^{i{\bf Q}\cdot{\bf  r_j}}
{\bf S_j}\right]^2\right\rangle,
\label{stag}
\ee
where ${\bf Q}=(\pi,\pi)$ and ${\bf S_j}$ is the electronic spin operator on site $j$
at position ${\bf r_j}$. In contrast to the behavior of the magnons, the fermions-to-pairs
ratio does not change considerably at moderate values of $t'$. Note that at $t'=0$ all the
holes appear as single fermions. This is a manifestation of the absence of pair-binding on
the single plaquette at $U=8t$.

We have found that the same behavior, both in terms of energetics and structure of the
ground state, persists across the entire range of geometries and doping levels which we
have studied. Therefore, we conclude that the initial rise of the pair-binding energy
for $t'<t'_{\rm max}$ is kinetic-energy driven. In this range most of the plaquettes
are in their half-filled, RVB-correlated ground-state. This type of background facilitates
the motion of bound pairs as compared to single holes. When $t'$ approaches $t'_{\rm max}$
the undoped background changes its nature and becomes more AFM. The gain in
kinetic energy associated with hole-pairing saturates and instead a gain in the potential
energy of unpaired holes sets in due to their interactions with the AFM magnons. This leads
to the decrease of the pair-binding energy.

Another correlation that we were able to establish is between
the maximum of the pair-binding energy and the position of the
single-hole ground state in momentum space. In both the
$4\times 4$ and $6\times 6$ periodic clusters the ground state
shifts from the $\Gamma-{\rm M}$ and symmetry related
directions of the Brillouin-zone to the zone-diagonals as $t'$
is increased through $t'_{\rm max}$, see Figs.
\ref{bindingCont1} and \ref{bindingCont2}. Specifically, exact
diagnonalization\cite{steve-exact-sup} of the $4\times 4$
cluster shows that the crystal momentum changes from $(0,\pi)$
and $(\pi,0)$ to $(0,0)$ and $(\pi,\pi)$ [CORE finds a similar
transition to $(\pi,\pi)$ but misses the $(0,0)$ state.] In the
$6\times 6$ cluster the shift is from $(0,\pm2\pi/3)$ and
$(\pm2\pi/3,0)$ to  $(\pm2\pi/3,\pm2\pi/3)$ [except for
$U=1-3t$ where in a narrow region above $t'_{\rm max}$ the
ground state is at $(0,0)$.]

It is known from quantum Monte-Carlo simulations that the single-hole ground state of
the {\it homogeneous} two-dimensional $t-J$ model resides at $(\pm\pi/2,\pm\pi/2)$.\cite{Assaad}
One may speculate whether this state is adiabatically connected to the ground state of the
inhomogeneous model for $t'>t_{\rm max}$. The answer to this question is beyond the
present study as it requires the diagonalization of larger clusters and the addition
of higher-energy plaquette fermions with plaquette momentum $(0,0)$ and $(\pi,\pi)$
to the effective Hilbert space. Regardless of this point, it seems that the transition
in the ground state momentum is a possible consequence of the maximum in $\Delta_{pb}$ rather
than its cause. We arrive at this conclusion based on the fact that in the $4\times 4$
cluster $\Delta_{pb}(3/16)$ exhibits a maximum of similar magnitude to that of
$\Delta_{pb}(1/16)$ while the 3-hole ground state is located at $(0,0)$ and $(\pi,\pi)$
over the entire parameter range.\cite{steve-exact-sup} In the $6\times 6$ cluster, on the
other hand, the maximum in $\Delta_{pb}(3/36)$ is accompanied by a change in the
3-hole ground state momentum, as depicted in Fig. \ref{bindingCont2}.

\subsection{Phase stiffness}
In the thermodynamic limit of a $d$-wave superconductor the pair-binding energy vanishes
as $\Delta_{pb}\sim 2\Delta_0 N^{-1/2}$, where $\Delta_0$ is the maximal value of the
superconducting gap.\cite{steve-exact} In our rather small clusters we can therefore
roughly estimate $\Delta_0\approx \Delta_{pb}/2$, which together with the $d$-wave BCS
gap relation $T_c=\Delta_0/2.14$, gives
\be
T_p=\frac{\Delta_{pb}}{4},
\label{Tp}
\ee
as a characteristic temperature at which pairs fall apart.

The actual $T_c$ may be smaller than $T_p$ if phase fluctuations are important.
To obtain an estimate for the phase-ordering temperature $T_\theta$
we calculate the ground-state phase stiffness defines as
\be
\rho_s=\left.\frac{1}{A}\frac{\partial^2E}{\partial\phi^2}\right|_{\phi=0}.
\label{rhos}
\ee
Here $E/A$ is the ground-state energy per unit area and $\phi$ is a phase twist
per bond in the $x$ direction.\cite{SWZ} Neglecting the suppression of the stiffness
due to thermal excitation of gapless nodal quasiparticles and using the relation
$T_c=0.89\rho_s$ for the two-dimensional $XY$ model we obtain the estimator
\be
T_\theta=\rho_s.
\label{Ttheta}
\ee

\begin{figure}[t]
\includegraphics[angle=0,width=\linewidth,clip=true]{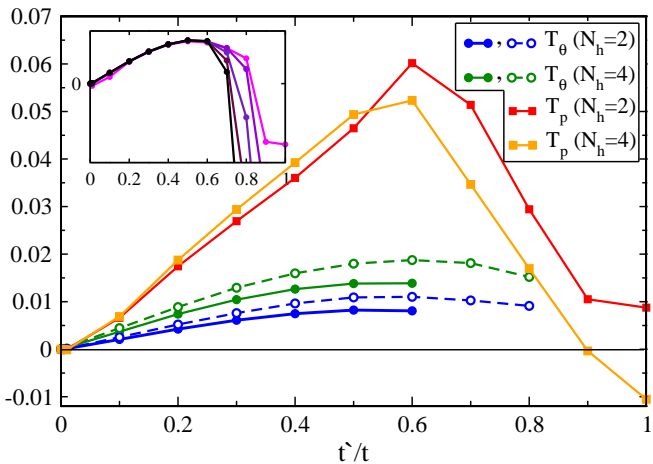}
\caption{The pairing scale $T_p$ and the phase coherence scale $T_\theta$ in the two-hole
and four-hole ground states of the $6\times 6$ periodic cluster at $U=8t$.
$T_\theta$ is shown for the cases where the phase twist is introduced at the bond level
(solid lines) and at the plaquette level (dashed lines). The inset depicts $T_\theta$
of the two-hole system as the phase twist
per bond $\phi$ is varied from $\pi/9$ (upper curve) to $\pi/72$ (lower curve).}
\label{stiff3x3}
\end{figure}

We have calculated $\rho_s$ in two ways. In the first the phase twist was introduced
into the Hamiltonian (\ref{H}) by changing $t_{ij}\rightarrow t_{ij}e^{i\phi/2}$ for
two nearest-neighbor sites in the $x$ direction. The effective CORE Hamiltonian for
the twisted system was then derived and diagonalized to obtain the $\phi$ dependence
of the ground state energy. In the second way the twist was introduced on the plaquette
level by modifying the couplings in the effective CORE Hamiltonian for the untwisted
model (\ref{H}). This was achieved via multiplication of a coupling between two
neighboring plaquettes in the $x$ direction that changes the number of holes on the
right plaquette by $\Delta n$, by $e^{i\phi\Delta n}$.

The phase-ordering temperature of the periodic $6\times 6$ cluster with two and four
holes is depicted in Fig. \ref{stiff3x3}. The two methods
yield similar results and they both encounter problems in the region $t'>t'_{\rm max}$.
The nature of the difficulty is demonstrated by the inset of Fig. \ref{stiff3x3},
showing $\rho_s$ as calculated from a discrete derivative of the ground state energy with
respect to a twist introduced at the bond level. When the derivative is calculated for
increasingly smaller values of $\phi$ the result does not converge for $t'>t'_{\rm max}$.
Rather, it becomes negative and diverges, indicating that the CORE ground-state energy
develops a cusp as function of $\phi$. A similar behavior is also found in the
$4\times 4$ periodic cluster and in the ladder systems. It occurs at lower values
of $t'$ for systems with odd number of holes. We take these findings as an indication
that CORE is unable to produce a reliable approximation for $\rho_s$ in the region
beyond the maximum in the pairing scale.

\begin{figure}[t]
\includegraphics[angle=0,width=\linewidth,clip=true]{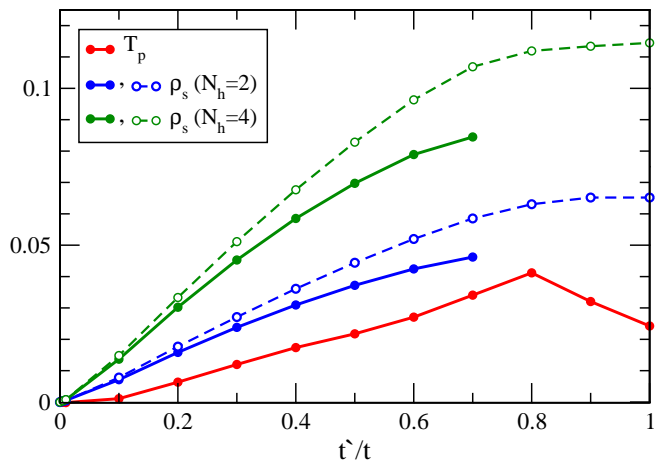}
\caption{The pairing scale $T_p$ and the phase stiffness of a $14\times 2$ ladder
in the two-hole and four-hole ground states at $U=8t$. $T_p$ is essentially
the same for the two hole doping levels. The solid (dashed) $\rho_s$ lines were
calculated by introducing the phase twist at the bond (plaquette) level.}
\label{stiffladder}
\end{figure}

In the range $t'<t'_{\max}$ the estimated phase-ordering temperature increases
monotonically with $t'$, but is consistently below the pairing scale. At $t'=t'_{\rm max}$
we find for the two-hole system $T_p/T_\theta\approx 6$. Increasing the doping to four
holes decreases the maximal $T_p$ slightly and increases $T_\theta$ by about 70\% leading to
$T_p/T_\theta(t'=t'_{\rm max})\approx 3$.
The same holds true for the $4\times 4$ cluster with two holes, which has
a similar hole density and $T_p/T_\theta(t'=t'_{\rm max})\approx 2$.
Such a behavior suggests that superconductivity in the lightly doped two-dimensional
checkerboard Hubbard model is governed by phase fluctuations.
In ladders our definition Eq. (\ref{rhos}) is equivalent to the phase stiffness along
the ladder ($v_cK_c$ in the effective Luttinger liquid
description of the system) divided by its width . As shown by Fig. \ref{stiffladder}
it is larger than the corresponding stiffness in the periodic clusters and grows with doping.
However, since the one-dimensional system can not order it does not provide a
phase ordering temperature similar to Eq. (\ref{Ttheta}).

\subsection{Pairing correlations}
Another diagnostic tool for the presence of superconductivity is the pair-field correlation
function. We have calculated the following equal-time correlator
\be
\label{delta-corr}
D_{\overline{ij},\overline{kl}}=\langle \Delta_{ij}^\dagger \Delta_{kl} \rangle ,
\ee
where $\overline{ij}$ denotes the bond between the nearest-neighbor sites $i$ and $j$,
and where the pair field on that bond is given by
\be
\label{delta-def}
\Delta_{ij}^\dagger=\frac1{\sqrt{2}} (c_{i\uparrow}^\dagger c_{j\downarrow}^\dagger
+c_{j\uparrow}^\dagger c_{i\downarrow}^\dagger).
\ee

\begin{figure}[t]
\includegraphics[angle=0,width=\linewidth,clip=true]{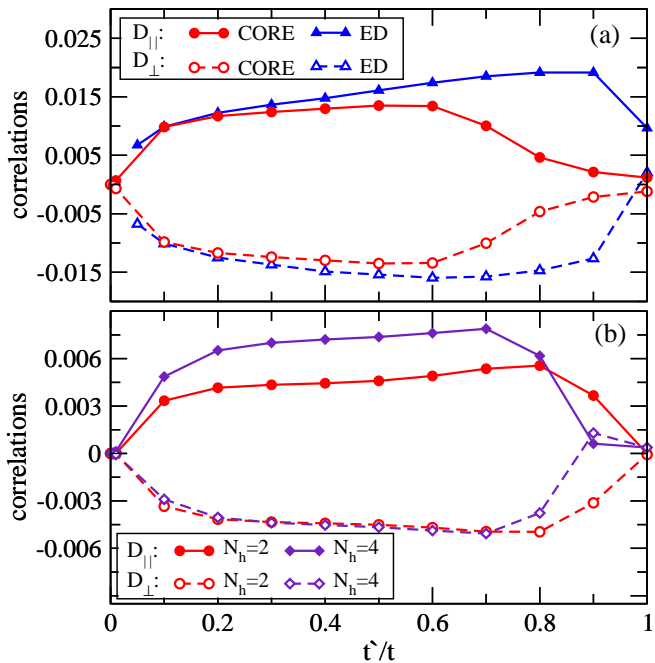}
\caption{Pair-field correlations at $U=8t$ in the ground state of
(a) the 2-hole doped $4\times 4$ periodic cluster, including a comparison to the exact
results of Ref. \onlinecite{steve-exact}, and (b) the 2-hole and 4-hole doped $6\times 6$
periodic cluster. $D_\perp$ and $D_\parallel$ are the correlations between the pair-field
on bond $\overline{ab}$ and the pair field on bonds $\overline{cd}$ and $\overline{ef}$,
respectively, as defined by Fig. \ref{model-fig}.}
\label{pairingCorr}
\end{figure}

Fig. \ref{pairingCorr} shows the results for the pair-field correlations between the two
most distant parallel $(D_\parallel)$ and perpendicular $(D_\perp)$ bonds on the periodic
clusters with $N_h=2$ and $N_h=4$. Similar results were also obtained for the ladder
systems. We find that $D_\parallel$ is positive and $D_\perp$ is negative,
consistent with $d$-wave pairing. The pairing correlations diminish in the limits
$t'/t\rightarrow 0$ and $t'/t\rightarrow 1$ but unlike the pair-binding energy and the
phase stiffness they are nearly independent of $t'$ in the range of moderate inhomogeneity
(from $t'=0.1t$ to $t'=0.6t$ $\Delta_{pb},\rho_s$ and $D$ change by a factor of 7.5,4.5,
and 1.5, respectively.) The magnitude of the correlations is small and comparable to
results of previous studies of Hubbard ladders\cite{2leg} and
Hubbard\cite{aimi} and $t-J$ periodic clusters.\cite{tj2d}

The behavior of $D$ suggests that pairing is very weak in the
systems that were studied. This conclusion is in apparent
contradiction with the large pair-binding energy found in the
same clusters. In addition, as we already noted, the
$t'$-dependence of the two quantities is very different. We
believe that the fault may lie in the specific form of the
pair-field, Eq. (\ref{delta-def}), that was used for
calculating the pairing correlations. It assumes a
pair-wavefunction which is strongly localized in space. This
may be wrong, as suggested by our results for the structure of
the ground-state. Fig. \ref{eigens2} clearly shows that most
holes are not bound into pairs on a single plaquette. This is
expected since for $U=8t$ the plaquette does not provide a
positive pair-binding energy. It seems, therefore, that
thinking about Cooper-pairing in such systems in terms of
real-space pairs occupying single bonds is a misleading
oversimplification. Most likely, the phenomenon is more
complicated and the pair wavefunction, while being much more
localized than its counterpart in conventional superconductors,
still possesses a non-trivial real-space structure.

\section{Conclusions}
\label{conclusions}

This study had a dual motivation. First, to explore the utility of the CORE approximation
as a method to investigate fermionic strongly correlated systems, and secondly to shed
additional light on the role of inhomogeneity in the physics of high-temperature
superconductivity.

As far as CORE is concerned, it is difficult to carry out the
original scheme of Morningstar and Weinstein\cite{CORE} who iteratively applied
the CORE method to obtain and analyze a fixed point Hamiltonian. In the case of
the Hubbard model there are simply too many couplings that are generated at each step.
One is, therefore, forced to apply CORE once and investigate the resulting effective
Hamiltonian either by means of a mean-field approximation\cite{AAcore}, or
via numerical diagonalization of finite clusters. The latter approach was
previously implemented in the study of spin systems\cite{Piekarewicz,capponi-spin}
and the $t-J$ model\cite{core-res,core-tj}, and is the one which we pursued.
As expected, when applied to the checkerboard Hubbard model range-2 CORE provides
results which are in good agreement with the available exact diagonalization results
in the limit of small $t'$.  In the moderate $t'$ regime the method may be considered
as semiquantitative and its validity in the uniform limit is questionable, particularly
in the case of small $U$. More precisely, this statement depends on the property that
one tries to calculate using the method. It seems that pairing is moderately local
such that range-2 CORE is able to capture its salient features already in small systems.
The establishment of phase coherence, on the other hand, is a more extended phenomenon,
for which the inclusion of longer range effective interactions and diagonalization of
larger clusters are needed. In this context we would like to note that signatures
associated with nodal quasiparticles of the putative $d$-wave Hubbard superconductor,
such as the suppression of the phase stiffness at low temperatures, are particularly
difficult to capture using range-2 CORE.\cite{AAcore}

Regarding the effects of inhomogeneity, our results demonstrate that plaquettization
of the Hubbard model may lead to a substantial enhancement of pairing. Optimal pairing
is achieved at an intermediate scale of inhomogeneity, which marks a crossover from a
region with pronounced RVB characteristics to one with stronger local AFM correlations.
The interactions of the doped holes with the spin background are the driving force
of the pairing process. One should bare in mind, however, that the Hubbard plaquette,
the building block of our model, is a special system. Its undoped ground state is a
quintessential RVB state and it provides a positive pair-binding energy in a wide
range of interaction strengths. Hence, it is interesting to ask whether a similar
enhancement occurs for other plane patterns, especially those constructed
from elementary clusters that do not exhibit pair binding. The possibility of
such an outcome gains support from the fact that in the checkerboard model
maximal pairing occurs at an interaction strength for which the pair-binding energy
on each individual plaquette is negative.

In the lightly doped clusters that we have studied superconductivity appears to be
controlled by phase fluctuations. Owing to the reasons outlined above and our inability
to carry out significant finite-size scaling it is difficult to estimate the phase
ordering temperature in the two-dimensional limit and determine whether
$T_c$ indeed achieves a maximum at an intermediate value of $t'$. $T_c$ enhancement
due to inhomogeneous pairing interaction was found in the attractive Hubbard
model\cite{optimal-BCS,optimal-s,optimal-QMC,optimal-BCS-dis,Mishra} and the
phase-ordering transition temperature is raised in the classical two-dimensional
$XY$ model with certain "framework" modulations of the phase
couplings.\cite{optimal-erica} We find it interesting to conclude by noting
that Fig. \ref{stiff3x3} hints at the possibility that a related inhomogeneity-induced
enhancement occurs in the model considered here as well.

\acknowledgments{We are most grateful to the authors of Ref. \onlinecite{steve-exact},
in particular to Wei-Feng Tsai, for providing us with the results of the exact
diagonalization of the $4\times 4$ Hubbard cluster. It is also our pleasure to thank
Sylvain Capponi for his valuable help in verifying our CORE calculations, and Erez Berg
for stimulating discussions.
This work was supported by the Israel Science Foundation (grants No. 538/08 and 459/09)
and by the United States - Israel Binational Science Foundation (grant No. 2008085).}

\appendix

\section{The CORE Hamiltonian}
\label{app:models}

The full CORE Hamiltonian includes all possible terms that satisfy the
symmetries of the problem, as detailed in Section \ref{models}. The resulting
45 effective couplings may be grouped in the following way
\begin{equation}
H=K_{bf}+K_{bf+t}+V_{bf}+V_{t}.
\end{equation}
The kinetic energy of the fermionic holes and the bosonic pairs is given by the
first two terms. $K_{bf}$ contains the contribution of hopping processes involving
only the charged degrees of freedom while $K_{bf+t}$ contains similar processes in
which the triplet of AFM magnons also participate. The interactions among the fermions
and pairs comprise $V_{bf}$. Their remaining interactions with the magnon triplet,
as well as couplings involving only the triplets, form the last group $V_t$.

In the following, $b_i^\dagger$, $t_{\sigma i}^\dagger$ and
$f_{\textbf{q} \sigma i}^\dagger$ create a hole pair, a magnon with spin component
$S_z=\sigma$ and a fermion with spin component $S_z=\sigma$ and plaquette momentum ${\bf q}$
at site $i$, respectively. Our choice to use a basis where the two fermions have a definite
plaquette momentum ${\bf q}=(0,\pi)$ or ${\bf q}=(\pi,0)$ results in different interaction
strengths between nearest neighbors in the $x$ direction compared to the $y$
direction. The notation $\langle i,j\rangle_\nu$ in the Hamiltonian below
stands for nearest neighbors in the $\nu=x,y$ direction and $(A_iB_j)_{S,\sigma}$ signifies that
the operators $A_i$ and $B_j$ are coupled into an operator of total spin $S$ and
spin component $S_z=\sigma$. Finally, summation over $S$ ,$\sigma$,
${\bf q}$, and $\nu$ indices is implied.

The 7 "bare" kinetic couplings include fermion and pair hopping, as
well as pair-fermion exchange and Andreev-like pair creation and disintegration.
\begin{eqnarray}
\label{Kbf}
\hspace{-0.9cm} K_{bf}&=& J_{b}  \sum_{\langle i,j\rangle}b_{i}^\dagger b_{j}\nonumber\\
&+& J_{f}^{\nu,{\bf q}}  \sum_{\langle i,j\rangle _{\nu}}f_{{\bf q}\sigma i}^\dagger
f_{{\bf q}\sigma j}\nonumber\\
&+& J_{bf}^{\nu,{\bf q}} \sum_{\langle i,j\rangle_\nu}b_i^\dagger f_{{\bf q}\sigma j}^\dagger
b_j f_{{\bf q}\sigma i}\nonumber\\
&+&J_{bff}^{\nu,{\bf q}}  \sum_{\langle i,j\rangle_{\nu}}\left[b_i^\dagger
f_{{\bf q}\uparrow i} f_{{\bf q}\downarrow j}+ b_i^\dagger f_{{\bf q}\uparrow j}
f_{{\bf q}\downarrow i}+{\rm H.c.}\right].
\end{eqnarray}
Note that since the Hamiltonian is symmetric under rotations and reflections
some of the couplings are related. For
example, $J_{f}^{x,{\bf q}}=J_{f}^{y,\overline{{\bf q}}}$, where
$\overline{{\bf q}}={\bf q}+(\pi,\pi) \mod 2\pi$.
These symmetries and the $d$-wave symmetry of the plaquette hole-pair state also imply
$J_{bff}^{x,{\bf q}}=-J_{bff}^{y,\overline{{\bf q}}}$.

The remaining 9 kinetic couplings are associated with magnon-assisted hopping processes
\begin{eqnarray}
\label{Kbft}
\hspace{-0.0cm}K_{bf+t}&=& J_{bt} \sum_{\langle i,j\rangle}b_i^\dagger t_{\sigma j}^\dagger b_j
t_{\sigma i} \nonumber\\
&+&J_{ft}^{S,\nu,{\bf q}} \sum_{\langle i,j\rangle_{\nu}}(t_i^\dagger
f_{{\bf q} j}^\dagger )_{S,\sigma} (t_j f_{{\bf q} i})_{S,\sigma}\nonumber\\
&+& J_{fft}^{\nu,{\bf q}} \sum_{\langle i,j\rangle_{\nu}}\left[(t_i^\dagger f_{{\bf q} j}
^\dagger )_{\frac{1}{2},\sigma} f_{\overline{{\bf q}}\sigma i}+{\rm H.c.}\right]\nonumber\\
&+& J_{bft}^{\nu,{\bf q}}  \sum_{\langle i,j\rangle_{\nu}}\left[b_i^\dagger t_{\sigma j}^\dagger
(f_{{\bf q} i} f_{\overline{{\bf q}} j})_{1,\sigma}+{\rm H.c.}\right].
\end{eqnarray}

The 16 fermion and pair on-site energies and interactions are
\begin{eqnarray}
V_{bf}&=& {\epsilon_{f}}_{\bf q}\sum_{i} f_{{\bf q}\sigma i}^\dagger
f_{{\bf q}\sigma i}+ \epsilon_b\sum_{i}  b_{i}^\dagger b_{i}\nonumber\\
&+& V_{b}  \sum_{\langle i,j\rangle}b_{i}^\dagger b_{j}^\dagger b_j b_i
+ V_{bf}^{\nu,{\bf q}} \sum_{\langle i,j\rangle_\nu}b_i^\dagger
f_{{\bf q}\sigma j}^\dagger f_{{\bf q}\sigma j} b_i\nonumber\\
&+& V1_{ff}^{S,\nu,{\bf q}} \sum_{\langle i,j\rangle_\nu}(f_{{\bf q} j}^\dagger
f_{{\bf q} i}^\dagger )_{S,\sigma} (f_{{\bf q} i}f_{{\bf q} j})_{S,\sigma}\nonumber\\
&+& V2_{ff}^{S} \sum_{\langle i,j\rangle_\nu}(f_{{\bf q} j}^\dagger
f_{{\bf q} i}^\dagger )_{S,\sigma} (f_{\overline{{\bf q}} i}
f_{\overline{{\bf q}} j})_{S,\sigma}\nonumber\\
&+& V3_{ff}^{S} \sum_{\langle i,j\rangle_\nu}(f_{{\bf q} j}^\dagger
f_{\overline{{\bf q}} i}^\dagger )_{S,\sigma} (f_{{\bf q} i}
f_{\overline{{\bf q}} j})_{S,\sigma}\nonumber\\
&+& V4_{ff}^{S} \sum_{\langle i,j\rangle_\nu}(f_{{\bf q} j}^\dagger
f_{\overline{{\bf q}} i}^\dagger )_{S,\sigma} (f_{\overline{{\bf q}} i}f_{{\bf q} j})_{S,\sigma}.
\end{eqnarray}
The fermion on-site energies depend on ${\bf q}$ in ladders where the symmetry between
the $x$ and $y$ directions is broken. The on-site energies on a plaquette depend on the
number of its nearest neighbors. Therefore, they may be position dependent in finite clusters
without periodic boundary conditions. This does not happen for the clusters that we have investigated.

The last group consists of 13 couplings involving the magnons. They include their on-site energy,
excitation amplitude from the vacuum, hopping matrix element and the strength of their mutual
interaction together with their interaction couplings to the fermions and bosons. We find the
coupling to the bosons to be very small.
\begin{eqnarray}
\label{Vt}
V_{t}&=& \epsilon_t \sum_{i} t_{\sigma i}^\dagger t_{\sigma i}
+J_{tt} \sum_{\langle i,j\rangle}\left[(t_{i}^\dagger t_{j}^\dagger)_0+{\rm H.c.}\right]\nonumber\\
&+& J_{t} \sum_{\langle i,j\rangle}t_{\sigma i}^\dagger t_{\sigma j}
+ V_{tt}^{S} \sum_{\langle i,j\rangle}(t_i^\dagger t_{j}^\dagger )_{S,\sigma} (t_{j} t_i)_{S,\sigma}
\nonumber\\
&+& V_{bt} \sum_{\langle i,j\rangle}b_i^\dagger t_{\sigma j}^\dagger t_{\sigma j} b_i \nonumber\\
&+& V_{ft}^{S,\nu,{\bf q}} \sum_{\langle i,j\rangle_{\nu}}(t_i^\dagger f_{{\bf q} j}^\dagger)
_{S,\sigma}(f_{{\bf q} j} t_i)_{S,\sigma}\nonumber\\
&+& V_{ft}^{\nu,{\bf q}}\sum_{\langle i,j\rangle_{\nu}}\left[(t_i^\dagger f_{{\bf q} j}^\dagger )
_{\frac{1}{2},\sigma} f_{\overline{{\bf q}}\sigma j}+{\rm H.c.}\right].
\end{eqnarray}

\end{document}